\documentclass[aps,preprint]{revtex4-1}%
\usepackage{amsfonts}
\usepackage{amsmath}
\usepackage{amssymb}
\usepackage{graphicx}%

\usepackage{url}
\usepackage[bookmarks,colorlinks,breaklinks]{hyperref}
\hypersetup{linkcolor=blue,citecolor=blue,filecolor=dullmagenta,urlcolor=blue}
\usepackage{bm,bbm}

\providecommand{\U}[1]{\protect\rule{.1in}{.1in}}

\begin{document}
\preprint{ }
\title{Generating infinite-dimensional algebras from loop algebras by expanding
Maurer Cartan forms.}
\author{R. Caroca$^{1,2}$, N. Merino$^{1}$, P. Salgado$^{1}$, O. Valdivia$^{1}$}
\affiliation{$^{1}$Departamento de F\'{\i}sica, Universidad de Concepci\'{o}n}
\affiliation{Casilla 160-C, Concepci\'{o}n, Chile. $\ \ \ \ \ $}
\affiliation{$^{2}$Departamento de Matem\'{a}tica y F\'{\i}sica Aplicadas, Universidad
Cat\'{o}lica de la Sant\'{\i}sima Concepci\'{o}n, Alonso de Rivera 2850
Concepci\'{o}n, Chile.}

\begin{abstract}
It is shown that the expansion methods developed in refs. \cite{azcarr} can be
generalized so that they permit to study the expansion of algebras of loops,
both when the compact finite-dimensional algebra and the algebra of loops have
a decomposition into two subspaces.

\end{abstract}
\maketitle

\section{Introduction}

Let $G(M)=G(S^{1})=Map(S^{1};G)$ be, the group of smooth mappings (loops)
$z\longrightarrow g(z)$ of the circle $S^{1}=\left\{  z\in C/\left\vert
z\right\vert =1\right\}  $ into a simple, compact and connected
finite-dimensional Lie group $G$. The group structure is defined by the
pointwise multiplication of functions $\left(  g%
\acute{}%
g\right)  (z)=g%
\acute{}%
(z)g(z).$ $Map(S^{1};G)$ is an infinite-dimensional group, the loop group
$LG,$ the elements of which can be \ represented by \cite{loop1},
\cite{azcarr1}
\begin{equation}
g(z)=e^{\alpha^{a}(z)T_{a}},\text{ \ }a=1,\cdot\cdot\cdot,r=\dim G \label{1}%
\end{equation}
where $T_{a}=-T_{a}^{\dag}$ are the generators of the finite-dimensional Lie
algebra $\mathcal{G}$, \ $\left[  T_{a},T_{b}\right]  =f_{ab}^{\text{ \ \ }%
c}T_{c}.$ For elements near the identity,
\begin{equation}
g(z)\simeq1+\alpha^{a}(z)T_{a}. \label{2}%
\end{equation}
Making a Laurent expansion of $\alpha^{a}(z)$ on the circle%
\begin{equation}
\alpha^{a}(z)=\sum_{n=-\infty}^{\infty}\alpha_{\text{ }-n}^{a}z^{n} \label{3}%
\end{equation}
expression (\ref{2}) reads
\begin{equation}
g(z)\simeq1+\sum_{n=-\infty}^{\infty}\alpha_{\text{ }-n}^{a}T_{a}z^{n}%
=1+\sum_{n=-\infty}^{\infty}\alpha_{\text{ }-n}^{a}T_{a}^{n}\text{ ,
\ \ \ }T_{a}^{n}\text{\ }\equiv T_{a}z^{n} \label{4}%
\end{equation}
where $T_{a}^{n}$ are the generators of the algebra $\widehat{\mathcal{G}%
}\equiv$ $\mathcal{G}(S^{1}).$ We may now write the commutation relations of
the Lie algebra in terms of the generators $T_{a}^{n}.$ The commutators of the
finite-dimensional $\mathcal{G}$ then imply
\begin{equation}
\left[  T_{\text{ }a}^{\text{ }m},T_{\text{ }b}^{\text{ }n}\right]
=f_{ab}^{\text{ \ }c}T_{c}^{\text{ }m+n}. \label{5}%
\end{equation}
Eqs. (\ref{5}) are the defining relations of the loop algebra associated with
$\mathcal{G}$, that is the algebra $\widehat{\mathcal{G}}=$ $\mathcal{LG}%
=Map(S^{1},\mathcal{G})$ of the loop group $LG.$ The original
finite-dimensional Lie algebra $\mathcal{G}$ is reproduced by the generators
$T_{\text{ }a}^{\text{ }0}$; they correspond to the generators of the group of
the constant maps $S^{1}\longrightarrow G$ whch is isomorphic to $G.$ With the
previous conventions, $T_{\text{ }a}^{\text{ }m\dagger}=-T_{\text{ }a}^{\text{
-}m}$ since, $z$ being of unit modulus, $z^{\ast}=z^{-1}.$

On the other hand, if $\{\omega^{a}(g)\}$, $a=1,...,r=dimG$, is the basis
determined by the (dual, left-invariant) Maurer--Cartan one-forms on $G$;
then, the Maurer-Cartan equations that characterize $\mathcal{G}$, in a way
dual to its Lie bracket description, are given by $d\omega^{c}=-\frac{1}%
{2}C_{ab}^{\text{ \ }c}\omega^{a}\wedge\omega^{b},$
\ \ \ $a,b,c=1,...,r=dimG.$

In direct analogy we can say that \ if $\{\omega^{a,n}\}$, $i=1,...,r=dimG$,
$n\in%
\mathbb{Z}
$ is the basis determined by the (dual, left-invariant) Maurer--Cartan
one-forms on $LG$; then, the corresponding Maurer-Cartan equations that
characterize the algebra $\widehat{\mathcal{G}}$, are given by%
\begin{equation}
d\omega^{c,l}=-\frac{1}{2}f_{(a,m)(b,n)}^{\text{ \ \ \ \ \ \ \ \ \ \ }%
(c,l)}\omega^{a,n}\wedge\omega^{b,m},\text{ \ \ \ }a,b,c=1,...,r=dimG;\text{
\ \ }l,m,n\in%
\mathbb{Z}
. \label{eins}%
\end{equation}%
\[
d\omega^{c,l}=-\frac{1}{2}\delta_{m+n}^{\text{ \ \ \ \ \ \ }l}f_{ab}^{\text{
\ \ }c}\omega^{a,n}\wedge\omega^{b,m},\text{ \ \ \ }a,b,c=1,...,r=dimG;\text{
\ \ }l,m,n\in%
\mathbb{Z}
.
\]

The purpose of this paper is to generalize the expansion procedures developed
in ref. \cite{azcarr} so that it permits to study the expansion of the
algebras of loops when both the compact finite-dimensional algebra
$\mathcal{G}$ \ and the loop algebra (which is an infinite-dimensional algebra
$\widehat{\mathcal{G}}$) have a decomposition into two subspaces $V_{0}\oplus
V_{1}.$

This article is organized as follow: In section $II$ we consider the rescaling
of the group parameters. In section $III$ we study $(i)$ \ the expansion of
the loop algebras when the compact finite-dimensional algebra $\mathcal{G}$
has a decomposition into two subspaces $\mathcal{G=}V_{0}\oplus V_{1}$ $(ii)$
\ the conditions under which the expanded algebra closes \ $(iii)$ the closure
of the expanded algebra when $V_{0}$ is a subalgebra. \ In section $IV$ we
study the expansion of the loop algebra (which is an infinite-dimensional
algebra $\widehat{\mathcal{G}}$), where this algebra $\widehat{\mathcal{G}}$
admits a decomposition $\widehat{\mathcal{G}}$ $=$ $V_{0}\oplus V_{1}.$
$\ $The expansion of $\widehat{\mathcal{G}}$ $=$ $V_{0}\oplus V_{1}$ when
$\left\{  V_{0},V_{1}\right\}  $ satisfy the condition of symmetric coset is
considered in section $V.$ Section $VI$ concludes the work with a brief comment.

\section{Rescaling of the group parameters \ and the expansion procedure}

Let $LG$ be the loop group, of local coordinates $g^{a}(z)$%
,\ $a=1,...,r=dimG.$ Let $\widehat{\mathcal{G}}$ be its algebra of basis
$\left\{  T_{a}^{n}\right\}  ,$ which may be realized by left-invariant
generators $T_{a}^{n}(g)$ on the group manifold. Let $\widehat{\mathcal{G}%
}^{\ast}$ be the coalgebra, and let $\{\omega^{a,n}(g)\},$ $i=1,...,r=dimG$,
$n\in%
\mathbb{Z}
$, \ be the basis $($dual, i.e., $\omega^{a,n}\left(  T_{b,m}\right)
\equiv\delta_{m}^{n}\delta_{b}^{a})$ determined by the Maurer-Cartan one-form
on $LG$. Then, when $\left[  T_{\text{ }a}^{\text{ }m},T_{\text{ }b}^{\text{
}n}\right]  =f_{ab}^{\text{ \ }c}T_{c}^{\text{ }m+n}$, \ the Maurer-Cartan
equations read
\begin{equation}
d\omega^{c,l}=-\frac{1}{2}f_{(a,m)(b,n)}^{\text{ \ \ \ \ \ \ \ \ \ \ }%
(c,l)}\omega^{a,n}\wedge\omega^{b,m},\text{ \ \ \ }a,b,c=1,...,r=dimG;\text{
\ \ }l,m,n\in%
\mathbb{Z}%
\end{equation}

\bigskip Let $\theta$ be the left-invariant canonical form on $LG$,
\begin{equation}
\theta(g)=g^{-1}dg=e^{-ig_{a,n}T^{a,n}}de^{ig_{a,m}T^{a,m}}\equiv\omega
^{a,n}T_{a,n},\text{ \ \ \ }a=1,...,r=dimG;\text{ \ \ }n\in%
\mathbb{Z}%
\end{equation}

\bigskip Since
\begin{equation}
e^{-A}de^{A}=dA+\frac{1}{2}\left[  dA,A\right]  +\frac{1}{3!}\left[  \left[
dA,A\right]  ,A\right]  +\frac{1}{4!}\left[  \left[  \left[  dA,A\right]
,A\right]  ,A\right]  +\cdot\cdot\cdot\cdot\cdot
\end{equation}
one obtains, for $A\equiv g_{a,n}T^{a,n},$ the expansion of $\theta(g)$ as
polynomials in the group coordinates $g^{a,n}:$
\begin{align}
\theta\left(  g\right)   &  =e^{-ig_{a,n}T^{a,n}}de^{ig_{a,m}T^{a,m}%
}\nonumber\\
&  =idg_{a_{1},n_{1}}T^{a_{1},n_{1}}+\frac{i^{2}}{2!}\left[  dg_{a_{2},n_{2}%
}T^{a_{2},n_{2}},g_{a_{3},n_{3}}T^{a_{3},n_{3}}\right] \nonumber\\
&  +\frac{i^{3}}{3!}[\left[  dg_{a_{2},n_{2}}T^{a_{2},n_{2}},g_{a_{3},n_{3}%
}T^{a_{3},n_{3}}\right]  ,g_{a_{4},n_{4}}T^{a_{4},n_{4}}]\nonumber\\
&  +\frac{i^{4}}{4!}[[\left[  dg_{a_{2},n_{2}}T^{a_{2},n_{2}},g_{a_{3},n_{3}%
}T^{a_{3},n_{3}}\right]  ,g_{a_{4},n_{4}}T^{a_{4},n_{4}}],g_{a_{5},n_{5}%
}T^{a_{5},n_{5}}]\nonumber\\
&  +\cdot\cdot\cdot\cdot\cdot\cdot\cdot\cdot\cdot\cdot\label{ec3.59}%
\end{align}
where the indices $a_{1},a_{2},a_{3}\cdot\cdot=1,2,...,\dim$ $\mathcal{G}$ ,
and $\ n_{i}\in%
\mathbb{Z}
$. \ Factoring the coordinates and their derivatives in the Lie brackets%
\begin{align}
\theta\left(  g\right)   &  =idg_{i_{1},n_{1}}T^{i_{1},n_{1}}+\frac{i^{2}}%
{2!}dg_{i_{2},n_{2}}g_{i_{3},n_{3}}\left[  T^{i_{2},n_{2}},T^{i_{3},n_{3}%
}\right] \nonumber\\
&  +\frac{i^{3}}{3!}dg_{i_{2},n_{2}}g_{i_{3},n_{3}}g_{i_{4},n_{4}}[\left[
T^{i_{2},n_{2}},T^{i_{3},n_{3}}\right]  ,T^{i_{4},n_{4}}]\nonumber\\
&  +\frac{i^{4}}{4!}dg_{i_{2},n_{2}}g_{i_{3},n_{3}}g_{i_{4},n_{4}}%
g_{i_{5},n_{5}}[[\left[  T^{i_{2},n_{2}},T^{i_{3},n_{3}}\right]
,T^{i_{4},n_{4}}],T^{i_{5},n_{5}}]\nonumber\\
&  +... \label{ec3.60}%
\end{align}
Using the commutation relation (\ref{5}) we have
\begin{align}
\left[  T^{a_{2},n_{2}},T^{a_{3},n_{3}}\right]   &  =if_{h_{1}}^{a_{2},a_{3}%
}T^{h_{1},n_{2}+n_{3}}\label{ec3.61}\\
\lbrack\left[  T^{a_{2},n_{2}},T^{a_{3},n_{3}}\right]  ,T^{a_{4},n_{4}}]  &
=i^{2}f_{h_{1}}^{a_{2},a_{3}}f_{h_{2}}^{h_{1},a_{4}}T^{h_{2},n_{2}+n_{3}%
+n_{4}}\label{ec3.62}\\
\lbrack\lbrack\left[  T^{i_{2},n_{2}},T^{i_{3},n_{3}}\right]  ,T^{i_{4},n_{4}%
}],T^{i_{5},n_{5}}]  &  =i^{3}f_{h_{1}}^{a_{2},a_{3}}f_{h_{2}}^{h_{1},a_{4}%
}f_{h_{3}}^{h_{2},a_{5}}T^{h_{3},n_{2}+n_{3}+n_{4}+n_{5}} \label{ec3.62'}%
\end{align}
so that (\ref{ec3.60}) takes the form%
\begin{align}
\theta\left(  g\right)   &  =idg_{a,n}T^{a,n}+\frac{i^{3}}{2!}dg_{a_{2},n_{2}%
}g_{a_{3},n_{3}}f_{a}^{a_{2},a_{3}}T^{a,n_{2}+n_{3}}\nonumber\\
&  +\frac{i^{5}}{3!}dg_{a_{2},n_{2}}g_{a_{3},n_{3}}g_{a_{4},n_{4}}f_{h_{1}%
}^{a_{2},a_{3}}f_{a}^{h_{1},a_{4}}T^{a,n_{2}+n_{3}+n_{4}}\nonumber\\
&  +\frac{i^{7}}{4!}dg_{a_{2},n_{2}}g_{a_{3},n_{3}}g_{a_{4},n_{4}}%
g_{a_{5},n_{5}}f_{h_{1}}^{a_{2},a_{3}}f_{h_{2}}^{h_{1},a_{4}}f_{a}%
^{h_{2},a_{5}}T^{a,n_{2}+n_{3}+n_{4}+n_{5}}\nonumber\\
&  +\cdot\cdot\cdot\cdot\cdot\cdot\cdot\cdot\label{ec3.65}%
\end{align}
expression that can be rewritten as%
\[
\theta\left(  g\right)  =[idg_{a,n}+\frac{i^{3}}{2!}\delta_{n}^{\left(
n_{2}+n_{3}\right)  }dg_{a_{2},n_{2}}g_{a_{3},n_{3}}f_{a}^{a_{2},a_{3}}%
\]%
\[
+\frac{i^{5}}{3!}\delta_{n}^{\left(  n_{2}+n_{3}+n_{4}\right)  }%
dg_{a_{2},n_{2}}g_{a_{3},n_{3}}g_{a_{4},n_{4}}f_{h_{1}}^{a_{2},a_{3}}%
f_{a}^{h_{1},i_{4}}%
\]%
\[
+\frac{i^{7}}{4!}\delta_{n}^{\left(  \alpha_{2}+\alpha_{3}+\alpha_{4}%
+\alpha_{5}\right)  }dg_{a_{2},n_{2}}g_{a_{3},n_{3}}g_{a_{4},n_{4}}%
g_{a_{5},n_{5}}f_{h_{1}}^{a_{2},a_{3}}f_{h_{2}}^{h_{1},i_{4}}f_{a}%
^{h_{2},a_{5}}%
\]%
\begin{equation}
+\cdot\cdot\cdot\cdot\cdot\cdot\cdot\cdot\cdot\cdot\cdot\cdot\cdot
]T^{a,n}\omega_{a,n} \label{ec3.67}%
\end{equation}

Therefore, the Maurer-Cartan 1-forms, $\omega_{a,n}(g)$, as a polynomial in
the coordinates of the group $g_{a,n}$ is given by
\[
\omega_{a,n}=idg_{a,n}+\frac{i^{3}}{2!}\delta_{n}^{\left(  n_{2}+n_{3}\right)
}dg_{a_{2},n_{2}}g_{a_{3},n_{3}}f_{a}^{a_{2},a_{3}}%
\]%
\[
+\frac{i^{5}}{3!}\delta_{n}^{\left(  n_{2}+n_{3}+n_{4}\right)  }%
dg_{a_{2},n_{2}}g_{a_{3},n_{3}}g_{a_{4},n_{4}}f_{h_{1}}^{a_{2},a_{3}}%
f_{a}^{h_{1},i_{4}}%
\]%
\begin{equation}
+\frac{i^{7}}{4!}\delta_{n}^{\left(  \alpha_{2}+\alpha_{3}+\alpha_{4}%
+\alpha_{5}\right)  }dg_{a_{2},n_{2}}g_{a_{3},n_{3}}g_{a_{4},n_{4}}%
g_{a_{5},n_{5}}f_{h_{1}}^{a_{2},a_{3}}f_{h_{2}}^{h_{1},i_{4}}f_{a}%
^{h_{2},a_{5}}+\cdot\cdot\cdot\cdot\cdot\cdot\cdot\label{ec3.68}%
\end{equation}
expression that can be rewritten as%
\[
\omega_{a,n}=idg_{a,n}++\frac{i^{3}}{2!}\delta_{n}^{\left(  n_{1}%
+n_{2}\right)  }dg_{a_{1},n_{1}}g_{a_{2},n_{2}}f_{a}^{a_{1},a_{2}}+%
{\displaystyle\sum\limits_{\beta=2}^{+\infty}}
\frac{i^{2\beta+1}}{\left(  \beta+1\right)  !}\delta_{n}^{\left(  n_{2}%
+n_{3}+\cdot\cdot\cdot\cdot+n_{\beta+1}\right)  }dg_{a_{1},n_{2}}%
g_{a_{2},n_{3}}%
\]%
\begin{equation}
\cdot\cdot\cdot\cdot\cdot\cdot\cdot\cdot\cdot\cdot g_{a_{\beta},n_{\beta+1}%
}g_{a_{\beta+1},n_{\beta+2}}f_{h_{1}}^{a_{1},a_{2}}f_{h_{2}}^{h_{1},a_{3}%
}...f_{h_{\beta-1}}^{h_{\beta-2},a_{\beta}}f_{a}^{h_{\beta-1},a_{\beta+1}}.
\label{3.69}%
\end{equation}

From (\ref{3.69}) we can see that the rescaling of some coordinates
$g_{i,\alpha}$
\begin{equation}
g_{a,n}\rightarrow\lambda g_{a,n} \label{ec3.92}%
\end{equation}
will generate an expansion of Maurer-Cartan 1-forms$\ \omega_{i,n}\left(
g,\lambda\right)  $ as a sum of 1-forms $\omega_{i,n}(g)$ on $LG$ multiplied
by the corresponding powers of $\lambda^{\alpha}$ of $\lambda$. This means
that the expansion (\ref{3.69}) exists and can be expressed as%
\begin{equation}
\omega_{i,n}=%
{\displaystyle\sum\limits_{\alpha=0}^{+\infty}}
\lambda^{\alpha}\omega_{i,n;\alpha}.
\end{equation}

\bigskip It should be noted that in the case $n=0$ and $n_{1}=n_{2}=\cdot
\cdot\cdot\cdot=n_{\beta+1}=0$ the equation (\ref{3.69}) takes the form%
\[
\omega_{a,0}=[i\delta_{a}^{a_{1}}+\frac{i^{3}}{2!}g_{a_{2},0}f_{a}%
^{a_{1},a_{2}}+%
{\displaystyle\sum\limits_{\beta=2}^{+\infty}}
\frac{i^{2\beta+1}}{\left(  \beta+1\right)  !}g_{a_{2},0}%
\]%
\begin{equation}
\cdot\cdot\cdot\cdot\cdot\cdot\cdot\cdot\cdot\cdot g_{a_{\beta},0}%
g_{a_{\beta+1},0}f_{h_{1}}^{a_{1},a_{2}}f_{h_{2}}^{h_{1},a_{3}}\cdot\cdot
\cdot\cdot f_{h_{\beta-1}}^{h_{\beta-2},a_{\beta}}f_{a}^{h_{\beta-1}%
,a_{\beta+1}}]dg_{a_{1},0}. \label{3.70}%
\end{equation}
That is, the equation (\ref{3.69}) reduces to the equation $(2.5)$ of ref.
\cite{azcarr}.

\section{Expansion of loop algebras \textbf{ }$\widehat{\mathcal{G}}$\textbf{
when }$\mathcal{G}=V_{0}\oplus V_{1}$}

In this section we consider the expansion of the loop algebras $\widehat
{\mathcal{G}}$\textbf{ \ }when the compact finite-dimensional algebra
$\mathcal{G}$ has a decomposition into two subspaces $\mathcal{G=}V_{0}\oplus
V_{1}$ $(ii)$ \ and we study the conditions under which the expanded algebra
closes. The case when $V_{0}$ is a subalgebra is also analized.

We consider the splitting of $\widehat{\mathcal{G}}^{\ast}$ into the sum of
two vector subspaces
\begin{equation}
\mathcal{G}^{\ast}=V_{0}^{\ast}\oplus V_{1}^{\ast},
\end{equation}
$V_{0}^{\ast}$ $,\ V_{1}^{\ast}$ being generated by the Maurer-Cartan forms
$\omega^{a_{0},n}\left(  g\right)  $, $\omega^{a_{1},n}\left(  g\right)  $ of
$\widehat{\mathcal{G}}^{\ast}$ with indices corresponding, respectively, to
the unmodified and modified parameters,%
\begin{equation}%
\begin{array}
[c]{ccc}%
g^{a_{0},n}\rightarrow g^{a_{0,n}}, & g^{a_{1},n}\rightarrow\lambda
g^{a_{1},n}, & a_{0}\left(  a_{1}\right)  =1,...,\dim V_{0}\left(  \dim
V_{1}\right)  ,\text{ }n\in%
\mathbb{Z}
.
\end{array}
\end{equation}
In general, the series of $\omega^{a_{0},n}(g,\lambda)$ $\in$ $V_{0}^{\ast}$,
$\omega^{a_{1},n}(g,\lambda)\in$ $V_{1}^{\ast}$ will involve all powers of
$\lambda$%
\begin{align}
\omega^{a_{p},n}\left(  g,\lambda\right)   &  =%
{\displaystyle\sum\limits_{\alpha=0}^{\infty}}
\lambda^{\alpha}\omega^{a_{p},n;\alpha}\left(  g\right) \nonumber\\
&  =\omega^{a_{p},n;0}\left(  g\right)  +\lambda\omega^{a_{p},n;1}\left(
g\right)  +\lambda^{2}\omega^{a_{p},n;2}\left(  g\right)  +.....,\text{
}p=0,1\text{ \ }%
\end{align}
where $\omega^{a_{p},n}\left(  g,1\right)  =\omega^{a_{p},n}\left(  g\right)
.$

With the above notation, the Maurer-Cartan equations (\ref{eins}) for
$\widehat{\mathcal{G}}$ can be rewritten as
\begin{equation}%
\begin{array}
[c]{cc}%
d\omega^{c_{s},l}=-\frac{1}{2}f_{a_{p},n\ \ b_{q},m}^{c_{s},l}\omega^{a_{p}%
,n}\omega^{b_{q},m} & (p,q,s=0,1)
\end{array}
\label{two}%
\end{equation}
where $a_{p},b_{q}=1,...,\dim V_{0}\left(  \dim V_{1}\right)  ;$ $l,n,m\in%
\mathbb{Z}
$ and where
\begin{equation}
\omega^{c_{s},l}=%
{\displaystyle\sum\limits_{\alpha=0}^{\infty}}
\lambda^{\alpha}\omega^{c_{s},l;\alpha}%
\end{equation}%
\begin{align}
\omega^{a_{p},n}  &  =%
{\displaystyle\sum\limits_{\alpha=0}^{\infty}}
\lambda^{\alpha}\omega^{a_{p},n;\alpha}\\
\omega^{b_{q},m}  &  =%
{\displaystyle\sum\limits_{\alpha=0}^{\infty}}
\lambda^{\alpha}\omega^{b_{q},m;\alpha}.
\end{align}
Introducing into the Maurer-Cartan (\ref{two}) we have
\begin{equation}%
{\displaystyle\sum\limits_{\alpha=0}^{\infty}}
\lambda^{\alpha}d\omega^{c_{s},l;\alpha}=-\frac{1}{2}c_{a_{p},n\ \ b_{q}%
,m}^{c_{s},l}%
{\displaystyle\sum\limits_{\alpha=0}^{\infty}}
\lambda^{\alpha}\omega^{a_{p},n;\alpha}%
{\displaystyle\sum\limits_{\beta=0}^{\infty}}
\lambda^{\beta}\omega^{b_{q},m;\beta}%
\end{equation}
and, using the eq. $(A.1)$ from Ref. \cite{azcarr}, the Maurer-Cartan
equations are expanded in powers of $\lambda:$
\begin{align}%
{\displaystyle\sum\limits_{\alpha=0}^{\infty}}
\lambda^{\alpha}d\omega^{c_{s},l;\alpha}  &  =-\frac{1}{2}c_{a_{p}%
,n\ \ b_{q},m}^{c_{s},l}%
{\displaystyle\sum\limits_{\alpha=0}^{\infty}}
\lambda^{\alpha}%
{\displaystyle\sum\limits_{\beta=0}^{\alpha}}
\omega^{a_{p},n;\beta}\omega^{b_{q},m;\alpha-\beta}\label{three}\\
&  =%
{\displaystyle\sum\limits_{\alpha=0}^{\infty}}
\lambda^{\alpha}\left(  -\frac{1}{2}c_{i_{p},n\ \ b_{q},m}^{c_{s},l}%
{\displaystyle\sum\limits_{\beta=0}^{\alpha}}
\omega^{a_{p},n;\beta}\omega^{b_{q},m;\alpha-\beta}\right)  .\nonumber
\end{align}
The equality of the two $\lambda$-polynomials in (\ref{three}) requires the
equality of the coefficients of equal power $\lambda^{\alpha}.$ This implies
that the coefficients one-forms $\omega^{a_{p},n_{2};\alpha}$ satisfy the
identities
\begin{equation}
d\omega^{c_{s},l;\alpha}=-\frac{1}{2}c_{a_{p},n\ \ b_{q},m}^{c_{s},l}%
{\displaystyle\sum\limits_{\beta=0}^{\alpha}}
\omega^{a_{p},n;\beta}\omega^{b_{q},m;\alpha-\beta}\text{, \ } \label{cuatro}%
\end{equation}
where $p,q,s=0,1;$ $a_{p},b_{q}=1,...,\dim V_{0}\left(  \dim V_{1}\right)  ;$
$l,n,m\in%
\mathbb{Z}
.$

We can rewrite (\ref{cuatro}) in the form
\begin{align}
d\omega^{c_{s},l;\alpha}  &  =-\frac{1}{2}C_{\left(  a_{p},n;\beta\right)
\ \ \left(  b_{q},m;\gamma\right)  }^{\left(  c_{s},l;\alpha\right)  }%
\omega^{a_{p},n;\beta}\omega^{b_{q},m;\gamma}\\
C_{\left(  a_{p},n;\beta\right)  \ \ \left(  b_{q},m;\gamma\right)  }^{\left(
c_{s},l;\alpha\right)  }  &  =\delta_{\beta+\gamma}^{\alpha}c_{a_{p}%
,n\ \ b_{q},m}^{c_{s},l}%
\end{align}
that is,
\begin{equation}
C_{\left(  a_{p},n;\beta\right)  \ \ \left(  b_{q},m;\gamma\right)  }^{\left(
c_{s},l;\alpha\right)  }=\left\{
\begin{array}
[c]{c}%
\begin{array}
[c]{ccc}%
0 & if & \beta+\gamma\neq\alpha
\end{array}
\\%
\begin{array}
[c]{ccc}%
c_{i_{p},n\ \ j_{q},m}^{k_{s},l} & if & \beta+\gamma=\alpha
\end{array}
\end{array}
\right.
\end{equation}
where $a_{p}$, $b_{q}$ , $c_{s}:$ $1,2,...,\dim\mathcal{G}$ , $l,n,m$ $\in$ $%
\mathbb{Z}
$ \ and $\alpha,\beta:$ $0,1,2,\cdot\cdot\cdot\cdot\cdot.$

Now we ask, under which conditions the 1-forms $\omega^{c_{0},l;\alpha_{0}}$,
$\omega^{c_{1},l;\alpha_{1}}$ generate new infinite dimensional algebras. The
answer is given by the following analysis: consider the one-forms%

\begin{equation}
\left\{  \omega^{a_{0},l;\alpha_{0}},\omega^{a_{1},l;\alpha_{1}}\right\}
=\left\{  \omega^{a_{0},l;0},\omega^{a_{0},l;1},...,\omega^{a_{0},l;N_{0}%
};\omega^{a_{1},l;0},\omega^{a_{1},l;1},...,\omega^{a_{1},l;N_{1}}\right\}
\label{mi9}%
\end{equation}
with $\alpha_{0}=0,...,N_{0}$, $\alpha_{1}=0,...,N_{1}$, $l\in$ $%
\mathbb{Z}
$. The conditions under which these forms generate new algebras are found by
demanding that the algebra generated by eq. (\ref{mi9}) is closed under the
exterior derivative $d$ and that the Jacobi identities for the new algebra are satisfied.

In fact, to find the conditions under which the algebra is closed, we write:%
\[
d\omega^{c_{s},l;\alpha}=-\frac{1}{2}c_{a_{p},n\ \ b_{q},m}^{c_{s},l}%
{\displaystyle\sum\limits_{\beta=0}^{\alpha}}
\omega^{a_{p},n;\beta}\omega^{b_{q},m;\alpha-\beta}%
\]%
\[
=-\frac{1}{2}c_{a_{0},n\ \ b_{0},m}^{c_{s},l}%
{\displaystyle\sum\limits_{\beta=0}^{\alpha}}
\omega^{a_{0},n;\beta}\omega^{b_{0},m;\alpha-\beta}-\frac{1}{2}c_{a_{0}%
,n\ \ b_{1},m}^{c_{s},l}%
{\displaystyle\sum\limits_{\beta=0}^{\alpha}}
\omega^{a_{0},n;\beta}\omega^{b_{1},m;\alpha-\beta}%
\]%
\begin{equation}
-\frac{1}{2}c_{a_{1},n\ \ b_{0},m}^{c_{s},l}%
{\displaystyle\sum\limits_{\beta=0}^{\alpha}}
\omega^{a_{1},n;\beta}\omega^{b_{0},m;\alpha-\beta}-\frac{1}{2}c_{a_{1}%
,n\ \ b_{1},m}^{c_{s},l}%
{\displaystyle\sum\limits_{\beta=0}^{\alpha}}
\omega^{a_{1},n;\beta}\omega^{b_{1},m;\alpha-\beta} \label{four}%
\end{equation}
\ \ which implies that
\begin{align}
d\omega^{c_{0},l;N_{0}}  &  =-\frac{1}{2}c_{a_{0},n\ \ b_{0},m}^{c_{0}%
,l}\left[  \omega^{a_{0},n;0}\omega^{b_{0},m;N_{0}}+...+\omega^{a_{0},n;N_{0}%
}\omega^{b_{0},m;0}\right] \label{m11}\\
&  -\frac{1}{2}c_{a_{0},n\ \ b_{1},m}^{c_{0},l}\left[  \omega^{a_{0}%
,n;0}\underset{(i)}{\omega^{b_{1},m;N_{0}}}+...+\omega^{a_{0},n;N_{0}}%
\omega^{b_{1},m;0}\right] \nonumber\\
&  -\frac{1}{2}c_{a_{1},n\ \ b_{0},m}^{c_{0},l}\left[  \omega^{a_{1}%
,n;0}\omega^{b_{0},m;N_{0}}+...+\underset{\left(  ii\right)  }{\omega
^{a_{1},n;N_{0}}}\omega^{b_{0},m;0}\right] \nonumber\\
&  -\frac{1}{2}c_{a_{1},n\ \ b_{1},m}^{c_{0},l}\left[  \omega^{a_{1}%
,n;0}\underset{\left(  iii\right)  }{\omega^{b_{1},m;N_{0}}}+...+\underset
{\left(  iv\right)  }{\omega^{a_{1},n;N_{0}}}\omega^{b_{1},m;0}\right]
\nonumber
\end{align}

\begin{align}
d\omega^{c_{1},l;N_{1}}  &  =-\frac{1}{2}c_{a_{0},n\ \ b_{0},m}^{c_{1}%
,l}\left[  \omega^{a_{0},n;0}\underset{\left(  v\right)  }{\omega
^{b_{0},m;N_{1}}}+...+\underset{\left(  vi\right)  }{\omega^{a_{0},n;N_{1}}%
}\omega^{b_{0},m;0}\right] \label{m12}\\
&  -\frac{1}{2}c_{a_{0},n\ \ b_{1},m}^{c_{1},l}\left[  \omega^{a_{0}%
,n;0}\omega^{b_{1},m;N_{1}}+...+\underset{\left(  vii\right)  }{\omega
^{a_{0},n;N_{1}}}\omega^{b_{1},m;0}\right] \nonumber\\
&  -\frac{1}{2}c_{a_{1},n\ \ b_{0},m}^{c_{1},l}\left[  \omega^{a_{1}%
,n;0}\underset{\left(  viii\right)  }{\omega^{b_{0},m;N_{1}}}+...+\omega
^{a_{1},n;N_{1}}\omega^{b_{0},m;0}\right] \nonumber\\
&  -\frac{1}{2}c_{a_{1},n\ \ b_{1},m}^{c_{1},l}\left[  \omega^{a_{1}%
,n;0}\omega^{b_{1},m;N_{1}}+...+\omega^{a_{1},n;N_{1}}\omega^{b_{1}%
,m;0}\right]  .\nonumber
\end{align}
Wherefrom we can see that the 1-forms $\omega^{b_{1},m;N_{0}}$ and
$\omega^{a_{1},n;N_{0}}$, corresponding to the terms identified by the symbols
$(i),(ii),(iii)$ and $(iv)$ in the equation \ref{m11}, belong to the base
(\ref{mi9}) if and only if
\begin{equation}
N_{0}\leq N_{1}. \label{cond1}%
\end{equation}
On the other hand, the 1-forms $\omega^{b_{0},m;N_{1}}$ and $\omega
^{a_{0},n;N_{1}},$ corresponding to the terms identified by the symbols
$(v),(vi),(vii)$ and $(viii)$ in the equaci\'{o}n (\ref{m12}), belong to the
base (\ref{mi9}) if and only if
\begin{equation}
N_{1}\leq N_{0}. \label{cond2}%
\end{equation}
From (\ref{cond1}-\ref{cond2}) it follows trivially that the conditions under
which the expanded algebra closes is
\begin{equation}
N_{0}=N_{1}. \label{mi12_2}%
\end{equation}

\section{\textbf{The case }$\widehat{\mathcal{G}}=V_{0}\oplus V_{1}$\textbf{
in which }$V_{0}$\textbf{ is a subalgebra }$L_{0}\subset$\textbf{ }%
$\widehat{\mathcal{G}}$\textbf{ }}

Let $\mathcal{G=}V_{0}\oplus V_{1}$, where now $V_{0}$ is a subalgebra
$\mathcal{L}_{0}$ of $\mathcal{G}$. From the commutation relation
\begin{equation}
\left[  T_{a,n},T_{b,m}\right]  =f_{ab}^{c}T_{c,n+m}=f_{a,n\ \ b,m}%
^{c,l}X_{c,l} \label{five}%
\end{equation}
$a_{p},b_{q}=1,...,\dim V_{0}\left(  \dim V_{1}\right)  ;$ $l,n,m\in%
\mathbb{Z}
.$ From (\ref{five}) we can see that $\mathcal{L}_{0}=\left\{  T_{a,0}%
\right\}  $ generates a subalgebra given by
\begin{equation}
\left[  T_{a,0},T_{b,0}\right]  =f_{ab}^{c}X_{c,0}=f_{a,0\ \ b,0}^{c,0}%
T_{c,0}. \label{six}%
\end{equation}
From (\ref{six}) we see that
\begin{equation}
f_{a,0\ \ b,0}^{c,n}=c_{ab}^{c}\delta_{0}^{n}=0\text{, para }n\neq0\text{,
\ }n\in%
\mathbb{Z}
. \label{s3}%
\end{equation}

Using (\ref{s3}) in the expansion%
\begin{align}
\omega^{a,n}\left(  g\right)   &  =[\delta_{\left(  b,m\right)  }^{\left(
a,n\right)  }+\frac{1}{2!}f_{b,m\ \ c,l}^{a,n}g^{c,l}\label{mcr4_KM}\\
&  +%
{\displaystyle\sum\limits_{r=2}^{\infty}}
\frac{1}{\left(  r+1\right)  !}f_{b,m\ \ c_{1},l_{1}}^{h_{1},p_{1}}%
f_{h_{1},p_{1}\ \ c_{2},l_{2}}^{h_{2},p_{2}}...\nonumber\\
&  ...f_{h_{r-2},p_{r-2}\ \ c_{r-1},l_{r-1}}^{h_{r-1},p_{r-1}}f_{h_{r-1}%
,p_{r-2}\ \ c_{r}l_{r}}^{a,n}g^{c_{1},l_{1}}g^{c_{2},l_{2}}...g^{c_{r-1}%
,l_{r-1}}g^{c_{r},l_{r}}]dg^{b,m}\nonumber
\end{align}
we find that under the rescaling
\begin{align}
g^{a,0}  &  \rightarrow g^{a,0},\ g^{a,n}\rightarrow\lambda g^{a,n}\text{
(}n\neq0\text{)},\ \nonumber\\
\left(  a,0\right)   &  =1,...,\dim V_{0}\nonumber\\
\left(  a,n\right)   &  =1,...,\dim V_{1}.\nonumber\\
V_{1}  &  =\left\{  T_{a,n}\right\}  \text{ with }n\neq0
\end{align}
the expansion of $\omega^{a,0}\left(  g,\lambda\right)  $ ($\omega
^{a,n}\left(  g,\lambda\right)  $ with $n\neq0$) starts with the power
$\lambda^{0}$ ($\lambda^{1}$). In fact, for $\omega^{a,0}\left(  g\right)  $
we have%
\begin{align}
\omega^{a,0}\left(  g\right)   &  =\left[  \delta_{\left(  b,n\right)
}^{\left(  a,0\right)  }+\frac{1}{2!}f_{b,n\ \ c,m}^{a,0}g^{c,m}+o\left(
g^{2}\right)  \right]  dg^{b,n}\nonumber\\
&  =dg^{a,0}+\frac{1}{2!}f_{b,n\ \ c,m}^{a,0}g^{c,m}dg^{b,n}+o\left(
g^{3}\right) \nonumber\\
&  =dg^{a,0}+\frac{1}{2!}\left(  f_{b,0\ \ c,0}^{a,0}g^{c,0}dg^{b,0}%
+f_{b,0\ \ c,n}^{a,0}g^{c,n}dg^{b,0}\right) \nonumber\\
&  +\frac{1}{2!}\left(  f_{b,n\ \ c,0}^{a,0}g^{c,0}dg^{b,n}+f_{b,n\ \ c,m}%
^{a,0}g^{c,m}dg^{b,n}\right)  +o\left(  g^{3}\right)
\end{align}
which implies that under the rescaling $g^{a,0}\rightarrow g^{a,0}%
,\ g^{a,n}\rightarrow\lambda g^{a,n}$ ($n\neq0$)$,$%
\begin{equation}
\omega^{a,0}\left(  g,\lambda\right)  =%
{\displaystyle\sum\limits_{\alpha=0}^{\infty}}
\lambda^{\alpha}\omega^{a,0;\alpha}\left(  g\right)  \label{sa3}%
\end{equation}
while for $\omega^{a,l}\left(  g\right)  $, with $l\neq0$, we have%

\begin{align}
\omega^{a,l}\left(  g\right)   &  =\left[  \delta_{\left(  b,n\right)
}^{\left(  a,l\right)  }+\frac{1}{2!}f_{b,n\ \ c,m}^{a,l}g^{c,m}+o\left(
g^{2}\right)  \right]  dg^{b,n}\\
&  =dg^{a,l}+\frac{1}{2!}f_{b,n\ \ k,m}^{a,l}g^{c,m}dg^{b,n}+o\left(
g^{3}\right) \nonumber\\
&  dg^{a,l}+\frac{1}{2!}(f_{b,0\ \ c,n}^{a,l}g^{c,n}dg^{b,0}+f_{b,n\ \ c,0}%
^{a,l}g^{c,0}dg^{b,n}+f_{b,n\ \ c,m}^{a,l}g^{c,m}dg^{b,n})+o\left(
g^{3}\right)  .\nonumber
\end{align}
Therefore the expansion of $\omega^{a,l}\left(  g,\lambda\right)  $ starts
with the power $\lambda^{1}$
\begin{equation}
\omega^{a,n}\left(  g,\lambda\right)  =%
{\displaystyle\sum\limits_{\alpha=1}^{\infty}}
\lambda^{\alpha}\omega^{a,n;\alpha}\left(  g\right)  . \label{saa4}%
\end{equation}
However, for computation purposes it is better to spread the sum from zero and
assume that $\omega^{a,n;0}=0$ for $n\neq0$. Thus we have that Eqs.
(\ref{sa3}-\ref{saa4}) can be summarized as: \ \ \
\begin{align}
\omega^{a,n}\left(  g,\lambda\right)   &  =%
{\displaystyle\sum\limits_{\alpha=0}^{\infty}}
\lambda^{\alpha}\omega^{a,n;\alpha}\left(  g\right) \label{z1}\\
\omega^{a,n;0}  &  =0\text{ for }n\neq0\text{.}\nonumber
\end{align}
Inserting (\ref{z1}) into the Maurer-Cartan equations $d\omega^{c,l}=-\frac
{1}{2}f_{a,n\ \ b,m}^{c,l}\omega^{a,n}\omega^{b,m},$ we have%
\begin{align}%
{\displaystyle\sum\limits_{\alpha=0}^{\infty}}
\lambda^{\alpha}d\omega^{c,l;\alpha}  &  =-\frac{1}{2}f_{a,n\ \ b,m}%
^{c,l}\left(
{\displaystyle\sum\limits_{\alpha=0}^{\infty}}
\lambda^{\alpha}\omega^{a,n;\alpha}\right)  \left(
{\displaystyle\sum\limits_{\beta=0}^{\infty}}
\lambda^{\beta}\omega^{b,m;\beta}\right) \\
&  =-\frac{1}{2}f_{a,n\ \ b,m}^{c,l}%
{\displaystyle\sum\limits_{\alpha=0}^{\infty}}
\lambda^{\alpha}%
{\displaystyle\sum\limits_{\beta=0}^{\alpha}}
\omega^{a,n;\beta}\omega^{b,m;\alpha-\beta}\nonumber\\
&  =%
{\displaystyle\sum\limits_{\alpha=0}^{\infty}}
\lambda^{\alpha}\left(  -\frac{1}{2}f_{a,n\ \ b,m}^{c,l}%
{\displaystyle\sum\limits_{\beta=0}^{\alpha}}
\omega^{a,n;\beta}\omega^{b,m;\alpha-\beta}\right)  .\nonumber
\end{align}
The equality of the coefficients of equal power $\lambda^{\alpha}$ leads to
the equation
\begin{align}
d\omega^{c,l;\alpha}  &  =-\frac{1}{2}f_{a,n\ \ b,m}^{c,l}%
{\displaystyle\sum\limits_{\beta=0}^{\alpha}}
\omega^{a,n;\beta}\omega^{b,m;\alpha-\beta}\label{z2}\\
&  =-\frac{1}{2}\delta_{n+m}^{l}f_{ab}^{c}%
{\displaystyle\sum\limits_{\beta=0}^{\alpha}}
\omega^{a,n;\beta}\omega^{a,m;\alpha-\beta}\nonumber
\end{align}
which can be rewritten as%
\begin{equation}
d\omega^{c,l;\alpha}=-\frac{1}{2}f_{\left(  a,n;\beta\right)  \ \ \left(
b,m;\gamma\right)  }^{\left(  c,l;\alpha\right)  }\omega^{a,n;\beta}%
\omega^{b,m;\gamma} \label{z3}%
\end{equation}
where
\begin{align}
f_{\left(  a,n;\beta\right)  \ \ \left(  b,m;\gamma\right)  }^{\left(
c,l;\alpha\right)  }  &  =\delta_{\beta+\gamma}^{\alpha}f_{a,n\ \ b,m}%
^{c,l}=\delta_{\beta+\gamma}^{\alpha}\delta_{n+m}^{l}f_{ab}^{c}\label{z4}\\
\omega^{a,n;0}  &  =0\text{ for }n\neq0\text{.}\nonumber
\end{align}

\subsection{\textbf{Analysis of }$\widehat{\mathcal{G}}\left(  N\right)
$\textbf{ for the cases N = 0,1, ...}}

Consider the form of equations (\ref{z2}).

For $\alpha=0$ \ we find:%
\begin{equation}
d\omega^{c,l;0}=-\frac{1}{2}\delta_{n+m}^{l}f_{ab}^{c}\omega^{a,n;0}%
\omega^{b,m;0}%
\end{equation}
but $\omega^{a,n;0}=0$ for $n\neq0$, we have
\begin{equation}
d\omega^{c,0;0}=-\frac{1}{2}f_{ab}^{c}\omega^{a,0;0}\omega^{b,0;0}.
\end{equation}
For $\alpha=1$ we find:%
\begin{align}
d\omega^{c,l;1}  &  =-\frac{1}{2}\delta_{n+m}^{l}f_{ab}^{c}%
{\displaystyle\sum\limits_{\beta=0}^{1}}
\omega^{a,n;\beta}\omega^{b,m;1-\beta}\\
&  =-\frac{1}{2}\delta_{n+m}^{l}f_{ab}^{c}\omega^{a,n;0}\omega^{b,m;1}%
-\frac{1}{2}\delta_{n+m}^{l}f_{ab}^{c}\omega^{a,n;1}\omega^{b,m;0}\nonumber\\
&  =-\frac{1}{2}\delta_{m}^{l}f_{ab}^{c}\omega^{a,0;0}\omega^{b,m;1}-\frac
{1}{2}\delta_{n}^{l}f_{ab}^{c}\omega^{a,n;1}\omega^{b,0;0}\nonumber\\
&  =-\frac{1}{2}f_{ab}^{c}\omega^{a,0;0}\omega^{b,l;1}-\frac{1}{2}f_{ab}%
^{c}\omega^{a,l;1}\omega^{b,0;0}\nonumber\\
&  =-\frac{1}{2}f_{ab}^{c}\omega^{a,0;0}\omega^{b,l;1}-\frac{1}{2}f_{ba}%
^{c}\omega^{b,l;1}\omega^{a,0;0}\nonumber\\
&  =-\frac{1}{2}f_{ab}^{c}\omega^{a,0;0}\omega^{b,l;1}+\frac{1}{2}f_{ab}%
^{c}\omega^{b,l;1}\omega^{a,0;0}\nonumber\\
&  =-\frac{1}{2}f_{ab}^{c}\omega^{a,0;0}\omega^{b,l;1}-\frac{1}{2}f_{ab}%
^{c}\omega^{a,0;0}\omega^{b,l;1}\nonumber\\
&  =-f_{ab}^{c}\omega^{a,0;0}\omega^{b,l;1}\nonumber
\end{align}%
\[
d\omega^{c,l;1}=-f_{ab}^{c}\omega^{a,0;0}\omega^{b,l;1}.
\]

In summary
\begin{align}
&
\begin{array}
[c]{cc}%
\alpha=0: & d\omega^{c,0;0}=-\frac{1}{2}f_{ab}^{c}\omega^{a,0;0}\omega
^{b,0;0};
\end{array}
\\
&
\begin{array}
[c]{cc}%
\alpha=1: & d\omega^{c,n;1}=-f_{ab}^{c}\omega^{a,0;0}\omega^{b,n;1};
\end{array}
\nonumber\\
&
\begin{array}
[c]{cc}%
\alpha\geq2: & d\omega^{c,l;\alpha}=-\frac{1}{2}\delta_{n+m}^{l}f_{ab}^{c}%
{\displaystyle\sum\limits_{\beta=0}^{\alpha}}
\omega^{a,n;\beta}\omega^{b,m;\alpha-\beta}%
\end{array}
.\nonumber
\end{align}
so that $\widehat{\mathcal{G}}\left(  0\right)  $ is given by
\begin{equation}
d\omega^{c,0;0}=-\frac{1}{2}f_{ab}^{c}\omega^{a,0;0}\omega^{b,0;0};
\end{equation}
and $\widehat{\mathcal{G}}\left(  1\right)  $ is given by
\begin{align}
d\omega^{c,0;0}  &  =-\frac{1}{2}f_{ab}^{c}\omega^{a,0;0}\omega^{b,0;0};\\
d\omega^{c,n;1}  &  =-f_{ab}^{c}\omega^{a,0;0}\omega^{b,n;1}.\nonumber
\end{align}

From the first equation we can see a non-trivial result: while for a
finite-dimensional Lie algebra $\mathcal{G}\left(  0\right)  =\mathcal{G}$,
for the loop \ algebra $\widehat{\mathcal{G}}\left(  0\right)  \neq
\widehat{\mathcal{G}}$ but $\widehat{\mathcal{G}}\left(  0\right)
=\mathcal{G}$ where $\mathcal{G}$ is the compact Lie algebra. \ 

\section{\textbf{The case }$\widehat{\mathcal{G}}=V_{0}\oplus V_{1}$\textbf{
in which }$V_{1}$\textbf{ is a symmetric coset }}

It is possible to consider the infinite-dimensional algebra as $\widehat
{\mathcal{G}}\mathcal{=}V_{0}\oplus V_{1}$ where $V_{0}$ is generated by the
infinite set of generators given by \ %

\begin{equation}
\left\{  ...,T_{a,-4},T_{a,-2},T_{a,0},T_{a,2},T_{a,4}...\right\}  \label{cs1}%
\end{equation}
and where $V_{1}$ is generated by%
\begin{equation}
\left\{  ...,T_{a,-3},T_{a,-1},T_{a,1},T_{a,3}...\right\}  . \label{cs2}%
\end{equation}
From the commutation relation
\begin{equation}
\left[  T_{a,n},T_{b,m}\right]  =f_{ab}^{c}T_{c,n+m} \label{cs3}%
\end{equation}
we clearly see that the condition for a symmetric coset is to satisfy: \ \
\begin{align}
\left[  V_{0},V_{0}\right]   &  \subset V_{0}\label{cs4}\\
\left[  V_{0},V_{1}\right]   &  \subset V_{1}\nonumber\\
\left[  V_{1},V_{1}\right]   &  \subset V_{0}.\nonumber
\end{align}
It is therefore interesting to study the expansion of the infinite-dimensional
algebra expanded with this choice of $V_{0}$ and $V_{1}$. For convenience we
distinguish the generators $T_{a,n}$ where the index $n$ is even from the case
when the index is odd. The most natural choice is to use a subscript zero
(one), $n_{0}\left(  n_{1}\right)  ,$ for even values (odd). Thus
(\ref{cs1}-\ref{cs3}) take the form:
\begin{equation}
\left\{  T_{a,n_{0}}\right\}  =\left\{  ...,T_{a,-4},T_{a,-2},T_{a,0}%
,T_{a,2},T_{a,4}...\right\}  , \label{cs5}%
\end{equation}%
\begin{equation}
\left\{  T_{a,n_{1}}\right\}  =\left\{  ...,T_{a,-3},T_{a,-1},T_{a,1}%
,T_{a,3}...\right\}  , \label{cs6}%
\end{equation}%
\begin{align}
\left[  T_{a,n_{0}},T_{b,m_{0}}\right]   &  =f_{ab}^{c}T_{c,n_{0}+m_{0}%
}=f_{a,n_{0}\ \ b,m_{0}}^{c,l_{0}}T_{c,l_{0}}\label{cs7}\\
\left[  T_{a,n_{0}},T_{b,m_{1}}\right]   &  =f_{ab}^{c}T_{c,n_{0}+m_{1}%
}=f_{a,n_{0}\ \ b,m_{1}}^{c,l_{1}}T_{c,l_{1}}\nonumber\\
\left[  T_{a,n_{1}},T_{b,m_{1}}\right]   &  =f_{ab}^{c}T_{c,n_{1}+m_{1}%
}=f_{a,n_{1}\ \ b,m_{1}}^{c,l_{0}}T_{c,l_{0}}.\nonumber
\end{align}
From where we see that the conditions of symmetric cosets for the structure
constants are given by
\begin{equation}
f_{a,n_{0}\ \ b,m_{0}}^{c,l_{1}}=f_{a,n_{0}\ \ b,m_{1}}^{c,l_{0}}%
=f_{a,n_{1}\ \ b,m_{1}}^{c,l_{1}}=0\text{.} \label{cs8}%
\end{equation}

The idea is: (a) to find the expansions of $\omega^{i,n_{0}}\left(
g,\lambda\right)  $ and $\omega^{i,n_{1}}\left(  g,\lambda\right)  ;$ (b) to
replace the expansions in the Maurer-Cartan equations and (c) to find the
conditions under which are generated new algebras.

To find the expansions of $\omega^{a,n_{0}}\left(  g,\lambda\right)  $ and
$\omega^{a,n_{1}}\left(  g,\lambda\right)  $ we must study the general
expansion of $\omega^{a,n_{0}}\left(  g\right)  $ and $\omega^{a,n_{1}}\left(
g\right)  $ in terms of the coordinates and then analyze the behavior under
the following rescaling:
\begin{align}
g^{a,n_{0}}  &  \rightarrow g^{a,n_{0}},\ g^{a,n_{1}}\rightarrow\lambda
g^{a,n_{1}}\label{cs9}\\
n_{0}  &  =...,-4,-2,0,2,4,...\nonumber\\
n_{1}  &  =...,-3,-1,1,3,....\nonumber
\end{align}
For $\omega^{a,n_{0}}\left(  g\right)  $ we find
\begin{align}
\omega^{a,n_{0}}\left(  g\right)   &  =\left[  \delta_{\left(  b,m\right)
}^{\left(  a,n_{0}\right)  }+\frac{1}{2!}f_{b,m\ \ c,l}^{a,n_{0}}%
g^{c,l}+o\left(  g^{2}\right)  \right]  dg^{b,m}\label{cs10}\\
&  =\delta_{\left(  b,m\right)  }^{\left(  a,n_{0}\right)  }dg^{b,m}+\frac
{1}{2!}f_{b,m\ \ c,l}^{a,n_{0}}g^{c,l}dg^{b,m}+o\left(  g^{3}\right)
\nonumber\\
&  =dg^{b,n_{0}}+\frac{1}{2!}f_{b,m\ \ c,l}^{a,n_{0}}g^{c,l}dg^{b,m}+o\left(
g^{3}\right) \nonumber\\
&  =dg^{b,n_{0}}+\frac{1}{2!}f_{b,m_{0}\ \ c,l_{0}}^{a,n_{0}}g^{c,l_{0}%
}dg^{b,m_{0}}+\frac{1}{2!}f_{b,m_{1}\ \ c,l_{1}}^{a,n_{0}}g^{c,l_{1}%
}dg^{b,m_{1}}+o\left(  g^{3}\right)  .\nonumber
\end{align}

Analyzing higher order terms we find that if you rescale the parameters as in
(\ref{cs9}), then $\omega^{a,n_{0}}\left(  g,\lambda\right)  $ contains only
even powers of $\lambda$. The proof is a direct generalization of the
procedure used in ref. \cite{azcarr}. For this it is useful to write the
condition (\ref{cs8}) as
\begin{equation}
f_{a,n_{p}\ \ b,m_{q}}^{c,l_{s}}=0,\text{ for }s\neq\left(  p+q\right)
\operatorname{mod}2. \label{cs11}%
\end{equation}

\bigskip Performing the same procedure for $\omega^{a,n_{1}}\left(
g,\lambda\right)  $ we find that appear in the expansion only odd powers of
$\lambda$. Thus we have%

\begin{align}
\omega^{a,n_{0}}\left(  g,\lambda\right)   &  =%
{\displaystyle\sum\limits_{\alpha=0}^{\infty}}
\lambda^{2\alpha}\omega^{a,n_{0};2\alpha}\left(  g\right) \label{cs12}\\
\omega^{a,n_{1}}\left(  g,\lambda\right)   &  =%
{\displaystyle\sum\limits_{\alpha=0}^{\infty}}
\lambda^{2\alpha+1}\omega^{a,n_{1};2\alpha+1}\left(  g\right) \nonumber
\end{align}
which can be written as
\begin{align}
\omega^{a,n_{p}}\left(  g,\lambda\right)   &  =\omega^{a,n_{\bar{\alpha}}%
}\left(  g,\lambda\right)  =\sum_{\alpha=0}^{\infty}\lambda^{\alpha}%
\omega^{a,n_{\bar{\alpha}};\alpha}\left(  g\right)  \text{;}\label{cs13}\\
\bar{\alpha}  &  =\alpha\operatorname{mod}2\text{, }p=0,1\text{.}\nonumber
\end{align}

Replacing (\ref{cs13}) in the Maurer-Cartan equations, we obtain the following
set of equations:%

\begin{equation}
d\omega^{c,l_{\bar{\alpha}};\alpha}=-\frac{1}{2}f_{\left(  a,n_{\bar{\beta}%
};\beta\right)  \left(  b,m_{\bar{\gamma}};\gamma\right)  }^{\left(
c,l_{\bar{\alpha}};\alpha\right)  }\omega^{a,n_{\bar{\beta}};\beta}%
\omega^{b,m_{\bar{\gamma}};\gamma} \label{cs14}%
\end{equation}
where
\begin{align}
f_{\left(  a,n_{\bar{\beta}};\beta\right)  \left(  b,m_{\bar{\gamma}}%
;\gamma\right)  }^{\left(  c,l_{\bar{\alpha}};\alpha\right)  }  &
=f_{a,n_{\bar{\beta}}\ \ b,m_{\bar{\gamma}}}^{c,l_{\bar{\alpha}}}\delta
_{\beta+\gamma}^{\alpha}\label{cs15}\\
\bar{\alpha}  &  =\alpha\operatorname{mod}2\text{, }\bar{\beta}=\beta
\operatorname{mod}2\text{, }\bar{\gamma}=\gamma\operatorname{mod}2.\nonumber
\end{align}

Performing the same procedure developed in ref. \cite{azcarr}, we find that
the expanded algebra (\ref{cs13}) closes when the coefficients of the
expansion are truncated at orders that satisfy the conditions%

\begin{align}
N_{1}  &  =N_{0}-1\text{, or}\\
N_{1}  &  =N_{0}+1.\nonumber
\end{align}

Now we consider some examples:

\subsubsection{The case in which $N_{1}=0$, $\widehat{\mathcal{G}}\left(
0,0\right)  :$}

If $N_{1}=0$ we have the trivial case $\widehat{\mathcal{G}}\left(
0,0\right)  =\widehat{\mathcal{G}}\left(  0\right)  $:%
\[
d\omega^{c,l_{0};0}=-\frac{1}{2}f_{\left(  a,n_{0};0\right)  \left(
b,m_{\bar{\gamma}};0\right)  }^{\left(  c,l_{0};0\right)  }\omega^{a,n_{0}%
;0}\omega^{b,m_{0};0}%
\]
which can be written as%
\begin{equation}
d\omega^{c,l_{0};0}=-\frac{1}{2}f_{a,n_{0}\ \ b,m_{0}}^{c,l_{0}}%
\omega^{a,n_{0};0}\omega^{b,m_{0};0}\text{.}%
\end{equation}
This means that, $\widehat{\mathcal{G}}\left(  0,0\right)  $ is the subalgebra
$\mathcal{L}_{0}=\left\{  T_{a,n_{0}}\right\}  $ of the original
infinite-dimensional algebra $\widehat{\mathcal{G}}$.

\subsubsection{The case in which $\widehat{\mathcal{G}}\left(  0,1\right)  $
is obtained as an In\"{o}n\"{u}-Wigner contraction of $\ \widehat{\mathcal{G}%
}:$}

Consider now the case $\widehat{\mathcal{G}}\left(  0,1\right)  $%
\begin{equation}
d\omega^{c,l_{0};0}=-\frac{1}{2}f_{a,n_{0}\ \ b,m_{0}}^{c,l_{0}}%
\omega^{a,n_{0};0}\omega^{b,m_{0};0}%
\end{equation}%
\begin{align}
d\omega^{c,l_{1};1}  &  =-\frac{1}{2}f_{\left(  a,n_{\bar{\beta}}%
;\beta\right)  \left(  b,m_{\bar{\gamma}};\gamma\right)  }^{\left(
c,l_{1};1\right)  }\omega^{a,n_{\bar{\beta}};\beta}\omega^{b,m_{\bar{\gamma}%
};\gamma}\\
&  =-\frac{1}{2}\left(  f_{\left(  a,n_{0};0\right)  \left(  b,m_{1};1\right)
}^{\left(  c,l_{1};1\right)  }\omega^{a,n_{0};0}\omega^{b,m_{1};1}+f_{\left(
a,n_{1};1\right)  \left(  b,m_{0};0\right)  }^{\left(  c,l_{1};1\right)
}\omega^{a,n_{1};1}\omega^{b,m_{0};0}\right) \nonumber\\
&  =-f_{\left(  a,n_{0};0\right)  \left(  b,m_{1};1\right)  }^{\left(
c,l_{1};1\right)  }\omega^{a,n_{0};0}\omega^{b,m_{1};1}\nonumber\\
&  =-f_{a,n_{0}\ \ b,m_{1}}^{c,l_{1}}\omega^{a,n_{0};0}\omega^{b,m_{1}%
;1}.\nonumber
\end{align}
This means that $\widehat{\mathcal{G}}\left(  0,1\right)  $ is given by%
\begin{align}
d\omega^{c,l_{0};0}  &  =-\frac{1}{2}f_{i,n_{0}\ \ j,m_{0}}^{c,l_{0}}%
\omega^{a,n_{0};0}\omega^{b,m_{0};0}\\
d\omega^{c,l_{1};1}  &  =-f_{a,n_{0}\ \ b,m_{1}}^{c,l_{1}}\omega^{a,n_{0}%
;0}\omega^{b,m_{1};1}%
\end{align}
i.e. $\widehat{\mathcal{G}}\left(  0,1\right)  $ corresponds to the
In\"{o}n\"{u}-Wigner contraction of $\widehat{\mathcal{G}}$ with respect to
$\mathcal{L}_{0}=\left\{  T_{a,n_{0}}\right\}  $: In fact, consider the
In\"{o}n\"{u}-Wigner contraction of%

\begin{align}
\left[  T_{a,n_{0}},T_{b,m_{0}}\right]   &  =f_{ab}^{c}T_{c,n_{0}+m_{0}%
}=f_{a,n_{0}\ \ b,m_{0}}^{c,l_{0}}T_{c,l_{0}}\\
\left[  T_{a,n_{0}},T_{b,m_{1}}\right]   &  =f_{ab}^{c}T_{c,n_{0}+m_{1}%
}=f_{a,n_{0}\ \ b,m_{1}}^{c,l_{1}}T_{c,l_{1}}\\
\left[  T_{a,n_{1}},T_{b,m_{1}}\right]   &  =f_{ab}^{c}T_{c,n_{1}+m_{1}%
}=f_{a,n_{1}\ \ b,m_{1}}^{c,l_{0}}T_{c,l_{0}}.
\end{align}

Rescaling the generators of the coset space $\widehat{\mathcal{G}}%
/\mathcal{L}_{0}$: $T_{a,n_{0}}=Y_{a,n_{0}}$, $T_{a,n_{1}}=\lambda Y_{a,n_{1}%
},$ we have%

\begin{align}
\left[  Y_{a,n_{0}},Y_{b,m_{0}}\right]   &  =f_{ab}^{c}Y_{c,n_{0}+m_{0}%
}=f_{a,n_{0}\ \ b,m_{0}}^{c,l_{0}}Y_{c,l_{0}}\\
\left[  Y_{a,n_{0}},Y_{b,m_{1}}\right]   &  =f_{ab}^{c}Y_{c,n_{0}+m_{1}%
}=f_{a,n_{0}\ \ b,m_{1}}^{c,l_{1}}Y_{c,l_{1}}\\
\left[  Y_{a,n_{1}},Y_{a,m_{1}}\right]   &  =\lambda^{-2}f_{ab}^{c}%
Y_{c,n_{1}+m_{1}}=\lambda^{-2}f_{a,n_{1}\ \ b,m_{1}}^{c,l_{0}}Y_{c,l_{0}}.
\end{align}

Taking the limit $\lambda\rightarrow\infty$ one finds%

\begin{align}
\left[  Y_{a,n_{0}},Y_{b,m_{0}}\right]   &  =f_{ab}^{c}Y_{c,n_{0}+m_{0}%
}=f_{a,n_{0}\ \ b,m_{0}}^{c,l_{0}}Y_{c,l_{0}}\\
\left[  Y_{a,n_{0}},Y_{b,m_{1}}\right]   &  =f_{ab}^{c}Y_{c,n_{0}+m_{1}%
}=f_{a,n_{0}\ \ b,m_{1}}^{c,l_{1}}Y_{c,l_{1}}\\
\left[  Y_{a,n_{1}},Y_{b,m_{1}}\right]   &  =0.
\end{align}

That is, the unique structure constants that are nonzero are $f_{a,n_{0}%
\ \ b,m_{0}}^{c,l_{0}}$ and $f_{a,n_{0}\ \ b,m_{1}}^{c,l_{1}}$. This means
that the equations
\begin{align}
d\omega^{c,l_{0};0}  &  =-\frac{1}{2}f_{a,n_{0}\ \ b,m_{0}}^{c,l_{0}}%
\omega^{a,n_{0};0}\omega^{b,m_{0};0}\\
d\omega^{c,l_{1};1}  &  =-f_{a,n_{0}\ \ b,m_{1}}^{c,l_{1}}\omega^{a,n_{0}%
;0}\omega^{b,m_{1};1}%
\end{align}
correspond to the In\"{o}n\"{u}-Wigner contraction of $\widehat{\mathcal{G}}$
with respect to $\mathcal{L}_{0}=\left\{  T_{a,n_{0}}\right\}  $. \ Notice
that the odd sector of the $\widehat{\mathcal{G}}$ algebra becomes abelian
after contraction.

\subsubsection{The Case $\widehat{\mathcal{G}}\left(  2,1\right)  $}

In this case we have,%

\begin{align}
d\omega^{a,l_{0};0}  &  =-\frac{1}{2}f_{a,n_{0}\ \ b,m_{0}}^{a,l_{0}}%
\omega^{a,n_{0};0}\omega^{b,m_{0};0}\\
d\omega^{c,l_{1};1}  &  =-f_{a,n_{0}\ \ b,m_{1}}^{c,l_{1}}\omega^{a,n_{0}%
;0}\omega^{b,m_{1};1},
\end{align}%
\begin{align}
d\omega^{c,l_{0};2}  &  =-\frac{1}{2}f_{\left(  a,n_{\bar{\beta}}%
;\beta\right)  \left(  b,m_{\bar{\gamma}};\gamma\right)  }^{\left(
c,l_{0};2\right)  }\omega^{a,n_{\bar{\beta}};\beta}\omega^{b,m_{\bar{\gamma}%
};\gamma}\nonumber\\
&  =-\frac{1}{2}(f_{\left(  a,n_{0};0\right)  \left(  b,m_{0};2\right)
}^{\left(  c,l_{0};2\right)  }\omega^{a,n_{0};0}\omega^{b,m_{0};2}+f_{\left(
a,n_{0};2\right)  \left(  b,m_{0};0\right)  }^{\left(  c,l_{0};2\right)
}\omega^{a,n_{0};2}\omega^{b,m_{0};0}+f_{\left(  a,n_{1};1\right)  \left(
b,m_{1};1\right)  }^{\left(  c,l_{0};2\right)  }\omega^{a,n_{1};1}%
\omega^{b,m_{1};1})\nonumber\\
&  =-f_{a,n_{0}\ \ b,m_{0}}^{c,l_{0}}\omega^{a,n_{0};0}\omega^{b,m_{0}%
;2}-\frac{1}{2}f_{a,n_{1}\ \ j,m_{1}}^{c,l_{0}}\omega^{a,n_{1};1}%
\omega^{b,m_{1};1}.
\end{align}
Thus $\widehat{\mathcal{G}}\left(  2,1\right)  $ is given by%
\begin{align}
d\omega^{c,l_{0};0}  &  =-\frac{1}{2}f_{a,n_{0}\ \ b,m_{0}}^{c,l_{0}}%
\omega^{a,n_{0};0}\omega^{b,m_{0};0}\\
d\omega^{c,l_{1};1}  &  =-f_{a,n_{0}\ \ b,m_{1}}^{c,l_{1}}\omega^{a,n_{0}%
;0}\omega^{b,m_{1};1}\\
d\omega^{c,l_{0};2}  &  =-f_{a,n_{0}\ \ b,m_{0}}^{c,l_{0}}\omega^{a,n_{0}%
;0}\omega^{b,m_{0};2}-\frac{1}{2}f_{a,n_{1}\ \ b,m_{1}}^{c,l_{0}}%
\omega^{a,n_{1};1}\omega^{b,m_{1};1}.
\end{align}
\bigskip and is generated by%
\begin{align}
&  \left\{  \omega^{a,n_{0};0};\omega^{a,n_{1};1},\omega^{a,n_{0};2}\right\}
\\
n_{0}  &  =...,-4,-2,0,2,4,...\nonumber\\
n_{1}  &  =...,-3,-1,1,3,....\nonumber
\end{align}

\section{\textbf{Comment}}

We have shown that the expansion methods developed in refs. \cite{azcarr} (see
also \cite{hatsuda}, \cite{deazcarraga2}) can be generalized so that they
permit to study the expansion of the algebras of loops both when the compact
finite-dimensional algebra $\mathcal{G}$ \ and the loop algebra (which is an
infinite-dimensional algebra $\widehat{\mathcal{G}}$) have a decomposition
into two subspaces $V_{0}\oplus V_{1}.$

This work was supported in part by Direcci\'{o}n de Investigaci\'{o}n,
Universidad de Concepci\'{o}n through Grant \# 210.011.053-1.0 and in part by
FONDECYT through Grants \# 1080530. Three of the authors (R.C, N.M and O.V)
were supported by grants from the Comisi\'{o}n Nacional de Investigaci\'{o}n
Cient\'{\i}fica y Tecnol\'{o}gica CONICYT and from the Universidad de
Concepci\'{o}n, Chile.


\begin{thebibliography}{9}                                                                                                %


\bibitem {azcarr}J.A. de Azcarraga, J.M. Izquierdo, M. Picon and O. Varela,
Nucl. Phys. B662 (2003) 185. $\left[  \text{arXiV:hep-th/0212347}\right]  ;$
Class. \& Quant. Grav. 21, S1375 (2004). $\left[  \text{arXiV:hep-th/0401033}%
\right]  $

\bibitem {loop1}P. Goddard and D. Olive, "Kac-Moody and Virasoro Algebras"
Int. J. Mod. Phys. A1 (1986) 303-414.

\bibitem {azcarr1}J.A. de Azcarraga, J.M. Izquierdo, "Lie grpoups, Lie
algebras, cohomology and some applications in physics". Cambridge University
Press 1998.

\bibitem {deazcarraga0}J.A. de Azc\'{a}rraga and J.C. P\'{e}rez Bueno, "Talk
given at 21st International Colloquium on Group Theoretical Methods in
Physics, Goslar, Germany, 15-20 July, 1996.$\left[
\text{arXiV:hep-th/9611221}\right]  .$

\bibitem {hatsuda}M. Hatsuda and M. Sakaguchi, Prog. Theor. Phys. 109, 853
(2003). $\left[  \text{arXiV:hep-th/0106114}\right]  .$

\bibitem {deazcarraga2}J.A. de Azcarraga, J.M. Izquierdo, M. Picon and O.
Varela, Int. J. Theor. Phys. 46 (2007) 2738. $\left[
\text{arXiV:hep-th/0703017v1}\right]  .$
\end{thebibliography}
\end{document}